\documentclass[10pt,a4paper]{article}
\usepackage{graphicx,amsmath,color}
\newcommand{\tr}{\textrm}
\newcommand{\sle}{$\textrm{SLE}_{\kappa}\,$}
\newcommand{\slek}{$\textrm{SLE}_{\kappa}$}
\newcommand{\pto}{\phi_{2}}
\newcommand{\ptho}{\phi_{3}}
\newcommand{\pl}{P_{\tr{left}}}
\newcommand{\pmm}{P_{\tr{middle}}}
\newcommand{\pr}{P_{\tr{right}}}
\newcommand{\p}{\partial}
\newcommand{\dd}{\textrm{d}}
\newcommand{\atan}{\arctan}

\newcommand{\es}{\frac{\epsilon}{16}}
\newcommand{\ee}{\frac{\epsilon}{8}}
\newcommand{\tes}{\frac{3\epsilon}{16}}
\newcommand{\h}{\frac{1}{2}}
\newcommand{\thh}{\frac{3}{2}}
\newcommand{\g}{\Gamma}

\begin{document}
\parindent 0mm
\parskip 6pt
\title{The scaling limit of two cluster boundaries in critical lattice models}
\author{Adam Gamsa and John Cardy\\
Rudolf Peierls Centre for Theoretical Physics\\
         1 Keble Road, Oxford OX1 3NP,
         U.K.}
\date{September 2005}
\maketitle
\begin{abstract}
The probability that a point is to one side of a curve in Schramm-Loewner evolution (\slek)
can be obtained alternatively using boundary conformal field theory (BCFT).
We extend the BCFT approach to treat two curves, forming, for example, the left
and right boundaries of a cluster. This proves to correspond to a generalisation to SLE$(\kappa,\rho)$, with $\rho=2$.
We derive the probabilities that a given point lies between two curves or to one side of both.
We find analytic solutions for the cases
$\kappa=0,2,4,8/3,8$. The result for $\kappa=6$ leads to predictions for the current distribution
at the plateau transition in the semiclassical approximation to the quantum Hall effect.
\end{abstract}
\section{Introduction}\label{Intro}
A large class of two-dimensional lattice models may be described
in terms of a gas of non-intersecting loops. Examples include the
boundaries of Ising spin clusters and percolation clusters, the
boundaries of the clusters in the random cluster representation of
the Potts model, the level lines of solid-on-solid models of the
roughening transition, as well as dilute self-avoiding walks and
polygons. These are in fact all special cases of the lattice
O$(n)$ model.

The continuum limit of these curves is conjectured to be both
scale and conformally invariant (a statement which may be made
precise and which has been proved in a few cases.) A description
of the whole ensemble of such curves is difficult, and instead it
is simpler to focus initially on a single curve connecting two
given points on the boundary of a simple connected domain, whose
existence is guaranteed by the boundary conditions. The
conformally invariant measure on such curves is conjectured to be
given by Schramm(stochastic)-Loewner Evolution (SLE) \cite{SchrammIsrael} with
parameter $\kappa$, where $n=-2\cos(4\pi/\kappa)$ with
$2\leq\kappa\leq8$. Many of the previously conjectured scaling
dimensions of the O$(n)$ model, as well as other properties such
as crossing formulae, have been derived using SLE \cite{CardyRev,Werner,LawlerBook}. The stochastic
description naturally leads to second-order linear differential
operators. Some of the scaling dimensions and correlation
functions are given in terms of suitable eigenvalues and
eigenfunctions of these.

An alternative description of the continuum limit of the critical
O$(n)$ model is in terms of conformal field theory (CFT). This
focusses on correlation functions of local scaling operators.
Within CFT there is a correspondence between such operators and
states in the radially quantised theory. These may be organised
into irreducible representations of the Virasoro algebra satisfied
by the generators $L_n$ of infinitesimal conformal
transformations.

A connection between these two pictures was made in
Refs.~\cite{BauerBernard,FriedrichWerner}. As was conjectured in 1984\cite{Cardybcc},
conditioning the CFT partition function on the existence of a
curve starting at a given boundary point is equivalent to the
insertion of a boundary operator\footnote{We use the notation
$\phi_n$ rather than the standard $\phi_{n,1}$ or $\phi_{1,n}$
because the parametrisation in terms of $\kappa$ does not
distinguish between the last two.} $\phi_2$ which corresponds to
a Virasoro representation with a level two null state:
$L_{-2}|\phi_2\rangle=\alpha L_{-1}^2|\phi_2\rangle$. As was also
shown in 1984\cite{BPZ}, this implies that the correlators of this
operator satisfy certain second-order linear differential
equations. These are the same as those coming from SLE$_\kappa$,
with the identification $\alpha=\kappa/4$.

However, CFT describes not only single curves but also many.
Indeed the correlator
\begin{equation}
\langle\phi_2(x_1)\phi_2(x_2)\ldots\phi_2(x_{2N})\ldots\rangle
\end{equation}
conditions the partition function on the existence of $N$ such
non-intersecting curves, hitting the boundary at the points
$\{x_j\}$. Such a correlation function satisfies $2N$ linear BPZ
equations, one for each $x_j$. Note that the order in which they
link up is not specified in the above. It may be made more precise
by assigning O$(n)$ labels to the $\phi_2$s: only operators with
the same label can then be connected by a curve. Alternatively, as
we shall discuss further, the different ways the curves link up
correspond to imposing different boundary conditions on the
differential equations.

\begin{figure}[ht]
\centering
\includegraphics[width=0.65\textwidth]{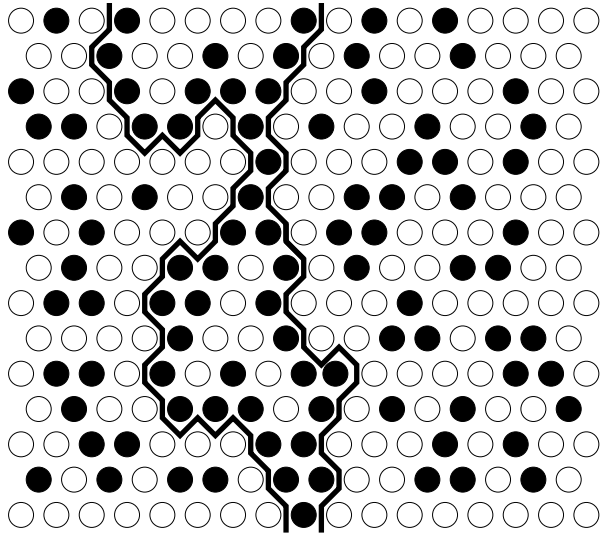}
\caption{An incipient infinite cluster containing a point on the
boundary in critical site percolation on the triangular lattice.
The scaling limit of the two curves forming its boundary is
described by the results of this paper.} \label{perc2curve}
\end{figure}

It is therefore straightforward in principle to derive within CFT
many different results relating to how $N$ such curves cross a
given domain. In practice, this becomes technically prohibitive.
In this paper we consider the simplest non-trivial case with
$N=2$. An interesting application of this for $\kappa=6$
(percolation) is the following problem, illustrated in
Fig~\ref{perc2curve}: consider critical site percolation in the
upper half plane, and suppose that all the sites on the real axis
are constrained to be white, except that at the origin, which is
black. Moreover this site is conditioned to be connected to
infinity by black sites, that is, it is part of an incipient
infinite black cluster. The boundaries of this cluster then define
two curves of the type we are considering.

One of the simplest SLE results for a single curve, due to
Schramm\cite{Schramm}, gives the probability that a given point
$\xi$ in the domain lies to the left (or right) of a single curve.
While this had not in fact previously been discussed within CFT,
its derivation from this point of view is
straightforward\cite{BBSLE}: the probability is given by the ratio
of conditional expectation values
\begin{equation}
\label{eqn:cond}
P(\xi)=\frac{\langle\Phi(\xi)\phi_2(x_1)\phi_2(x_2)\rangle}
{\langle\phi_2(x_1)\phi_2(x_2)\rangle}\,,
\end{equation}
where $\Phi(\xi)$ is the indicator function that the curve passes
to the left(right) of $\xi$. [This behaves to all intents and
purposes like a local operator with scaling dimension zero.]
Specialising (without loss of generality, because of conformal
invariance) to the case when the domain is the upper half plane
and $x_2\to\infty$, the denominator becomes trivial and the
numerator satisfies a BPZ equation with respect to $x_1$.

The generalisation of Eq.~(\ref{eqn:cond}) to $N$ curves is
straightforward. For example, for two curves in the upper half
plane conditioned to connect $(x_1,\infty)$ and $(x_2,\infty)$
\begin{equation}
\label{2curves}
P(\xi)=\frac{\langle\Phi(\xi)\phi_2(x_1)\phi_2(x_2)\phi_2(\infty)
\phi_2(\infty)\rangle}
{\langle\phi_2(x_1)\phi_2(x_2)\phi_2(\infty)
\phi_2(\infty)\rangle}\,,
\end{equation}
where now $\Phi(\xi)$ is any of the indicator functions that $\xi$
lies to the left, or right, of both curves, or between them. Both
numerator and denominator satisfy 2nd-order BPZ equations with
respect to $x_1$ and $x_2$: the boundary conditions pick out which
case is being computed.

These coupled partial differential equations are already too
difficult to solve in closed form. However, they simplify in the
limit when $\delta=x_2-x_1\ll|\xi|$, by a well-known property of
CFT called the fusion rules. These state, roughly speaking, that
in this limit the operator product $\phi_2(x_1)\phi_2(x_2)$ may be
written
\begin{equation}
\label{ope}
\phi_2(x_1)\cdot\phi_2(x_2)=\delta^{\alpha_1}\phi_1(x_1)+
\delta^{\alpha_2}\phi_3(x_1)\,,
\end{equation}
where the correlators of $\phi_1$ and $\phi_3$ satisfy
respectively first and third order equations. In fact $\phi_3$
corresponds to a Virasoro representation with a null state at
level 3. The null state condition may be computed explicitly, and
hence the third-order equation. (This could also be obtained
directly from the two 2nd order equations.) We also argue that if
the curves are conditioned to go to infinity (rather than there
being a single curve connecting $x_1$ and $x_2$) this picks out
the second term in Eq.~(\ref{ope}). The resulting correlation
functions now depend only on $\arg(\xi-x_1)$ and may be written as
indefinite integrals of ordinary hypergeometric functions.

The layout of this paper is as follows. In the next section we
recall the derivation of Schramm's formula for a single curve
using both SLE and CFT. Sec.~\ref{DiffEqn2Curves} contains the main body of
the CFT calculation for two curves. In Sec.~\ref{results} we present
the results, in graphical form for $\kappa=6,8/3$. For some other
values of $\kappa$ they may be expressed in terms of elementary
functions. The limiting cases $\kappa=0$ and $\kappa=8$ are
interesting.

In Sec.~\ref{qhall} we apply the results for $\kappa=6$ to the
semi-classical limit of the quantum Hall plateau transition, where
electrons in a strong magnetic field move in random scalar
potential. If Coulomb interactions may be neglected, the guiding
centres of the electrons approximately follow the level lines of
the potential, which, in the scaling limit, are percolation
cluster boundaries. At the critical point the half-plane geometry
may be conformally transformed into a long strip, and our
calculation gives information about the paths followed by the
conduction electrons, and hence the mean current density across
the strip.

Finally in Sec.~\ref{RevEng} we discuss the equivalence between
the BPZ equations of CFT and a postulated generalisation of SLE to
multiple curves. The growth of a single curve, conditioned on the
existence of the others, is described by a variant of SLE known as
SLE$(\kappa,\rho)$ (with $\rho=2$), as asserted in several recent
studies of multiple curves
\cite{Dubedat,BauerBernardKytola,Kytola}. The CFT equations also
suggest \cite{CardyDyson,CardyCal} that it is possible to describe
the joint measure on all the curves by a `multiple SLE' in which
they are grown simultaneously, as shown precisely in
\cite{BauerBernardKytola}.

\section{Schramm's formula}
In this section we review the computation of the probability that
a given point $\xi$ lies to the left(right) of a single curve,
from the points of view of both SLE and CFT.
\subsection{CFT method}
As discussed in Sec.~\ref{Intro}, the probability is given as the ratio of
correlators in Eq.~(\ref{eqn:cond}). We take $x_1\to x$ and $x_2$
large (and eventually to infinity). Consider the effect of the
infinitesimal conformal transformation $z\to z+\epsilon/(z-x)$,
which is implemented by inserting into each correlator a factor
\begin{equation}
\int_C\frac{\epsilon\, T(z)}{z-x}\frac{dz}{2\pi i}\,,
\end{equation}
where $C$ is a contour surrounding $x$ (but not $\xi$ or $x_2$),
together with its reflection in the real axis. This may be
evaluated in two ways: by shrinking the contour around $z=x$ and
using the fact that the $O\big((z-x)^0\big)$ term in the operator
product expansion of $T(z)$ with $\phi_2(x)$ is (by definition)
$L_{-2}\phi_2(x)=(\kappa/4)L_{-1}^2\phi_2(x)=(\kappa/4)\partial_x^2\phi_2(x)$; or
by wrapping the contour around $\xi$ and $x_2$. The effect on
$\Phi(\xi)$ is just to shift $\xi$, while the effect on
$\phi_2(x_2)$ is negligible as $x_2\to\infty$.

Equating these two ways of evaluating the insertion gives, as
$x_2\to\infty$, the BPZ equation
\begin{equation}
\label{BPZ1}
 \left(\frac\kappa 4\frac{\partial^2}{\partial x^2}
+2{\rm Re}\Big[
\frac1{\xi-x}\frac\partial{\partial\xi}\Big]\right)P(\xi,x)=0\,,
\end{equation}
where $P(\xi,x)$ is either probability. Since these only depend on
the angle which $\xi-x$ makes with the axis, or equivalently the
variable $t\equiv{\rm Re}(\xi-x)/{\rm Im}\xi$, this partial
differential equation reduces to an ordinary one, of Riemann type.
The boundary conditions are $P_{\rm left}\to0$ as $t\to+\infty$
and $P_{\rm left}\to1$ as $t\to-\infty$. The equation itself allows
these two possible asymptotics -- an explicit calculation
determines the exponents as $|t|\to\infty$ to be $|t|^{-\gamma}$
with $\gamma=0$ or $\gamma=\tilde x_2=(8-\kappa)/\kappa$. The
latter is the boundary 2-leg exponent: it arises as $t\to+\infty$
since the point $\xi$ traps the curve against the real axis so that, in
effect, two mutually avoiding curves of the O$(n)$ model emerge
from that point.

These boundary conditions then fix the solution to be
\begin{equation}\label{fix}
P_{\rm
left}=\h-\frac{\g(\frac{4}{\kappa})}{\sqrt{\pi}\g(\frac{8-\kappa}{2\kappa})}t\,_{2}F_{1}
(\h,\frac{4}{\kappa};\frac{3}{2};-t^2) \,.
\end{equation}
Note that, because $P_{\rm left}+P_{\rm right}=1$ is
a solution to the equation, any other, including Eq.~(\ref{fix}), may be
written as a quadrature of an elementary function.
\subsection{SLE method}
We summarise the theoretical physicist's version\cite{CardyRev,BauerBernard} of
Schramm's original argument. In SLE, the curve in the upper half
plane from a point $a_0$ on the real axis to infinity is
considered as being grown dynamically, introducing a fictitious
time variable $t$ (obviously distinct from the variable $t$
defined above.) Let $K_t$ be the set consisting of the curve as
grown up to time $t$ (as well as all points enclosed by the curve
and between the curve and the real axis) so that the complement of
this set in the upper half plane is simply connected. Let $g_t(z)$
be the (unique) conformal mapping of this complement to the whole
upper half plane, normalised such that $g_t(z)=z+O(1/z)$ as
$z\to\infty$. The coefficient of the $O(1/z)$ term is increasing
with $t$, so `time' can be reparametrised so this coefficient is
exactly $2t$. The image of the growing tip of the curve under
$g_t$ is a point $a_t$ on the real axis. Loewner showed that the
time-evolution of $g_t$ satisfies
\begin{equation}
\label{Loewnereqn} \frac{dg_t(z)}{dt}=\frac2{g_t(z)-a_t}\,.
\end{equation}
Any suitably continuous function $a_t$ generates a curve.
Schramm\cite{SchrammIsrael} showed that if this process is to generate a
conformally invariant measure on curves, the only possibility is
$a_t=\sqrt\kappa B_t +a_0$, with $B_t$ being a standard Brownian
motion.

Now consider the problem at hand, with a curve $\gamma$ connecting
$a_0$ to $\infty$, and a given point $\xi$ away from $a_0$. Evolve
the SLE for an infinitesimal time $dt$. The function $g_{dt}$ will
erase a short initial segment, and map the remainder of $\gamma$
into its image $\gamma'$, which, however, by conformal invariance,
will have the same measure as SLE started from
$a_{dt}=a_0+\sqrt\kappa dB_t$. At the same time, $\xi\to
\xi'=\xi+2dt/(\xi-a_0)$. Moreover, $\gamma'$ lies to the
left(right) of $\xi'$ iff $\gamma$ lies to the left(right) of
$\xi$. Therefore
\begin{equation}
P\big(\xi;a_0\big)= \langle P\big(\xi+2dt/(\xi-a_0),a_0
+\sqrt\kappa dB_t\big)\rangle\,,
\end{equation}
where the average $\langle\ldots\rangle$\ is over all realisations
of Brownian motion $dB_t$ up to time $dt$. Taylor expanding, using
$\langle dB_t\rangle=0$ and $\langle(dB_t)^2\rangle=dt$, and
equating the coefficient of $dt$ to zero then gives exactly the
CFT Eq.~(\ref{BPZ1}) if we set $a_0\to x$.

\section{The differential equation for two curves}\label{DiffEqn2Curves}
According to CFT the probability
that a given point $\xi$ lies to the left, between or to the right
of two curves starting at points $x_1,x_2$ on the real axis is
given by a ratio of correlators as in Eq.~(\ref{2curves}). The numerator
and denominator each satisfy BPZ equations with respect to both
$x_1$ and $x_2$. For general values, these are discussed further
in Sec.~\ref{RevEng}. However, explicit analytic progress is only
feasible in the limit when $\delta=x_2-x_1\to0$, and we now treat
this using established properties of CFT.

The operator product expansion of two $\phi_2$ operators is
constrained by the fusion rules to have the form in Eq.~(\ref{ope}). This
means that every solution of the coupled second order BPZ
equations may be written in this limit as a linear combination of
functions $\delta^{\alpha_j}F_j(\xi,x,\delta)$, with $j=1,3$ and
$x=(x_1+x_2)/2$, and the functions $F_j$ having a regular power
series expansion in $\delta$. The values of the $\alpha_j$ are
determined by the differential equations to be
$\alpha_1=-2h_2=-(6-\kappa)/\kappa$ and
$\alpha_3=h_3-2h_2=2/\kappa$.

In general, the dominant behaviour as $\delta\to0$ is given by
$\alpha_1$, corresponding to the first term $\phi_1$ in the OPE,
Eq.~(\ref{ope}). This is just the identity operator, and it is
straightforward to see that the corresponding solution $F_1$ is in
fact a constant. The physical interpretation of this is that
$x_1,x_2$ are overwhelming likely to be the end-points of the \em
same \em curve as $\delta\to0$, which makes a very small excursion
into the upper half plane. It has no effect on conditional
probabilities of events further away. In order to condition the
two curves each to go to infinity, we must therefore impose the
condition that this term is absent in the solution.

This leaves the term in the OPE coupling to the
$\phi_3$ operator. The leading behaviour of the probability function, Eq.~(\ref{2curves}), in
this limit is therefore given by the ratio
\begin{equation}\label{SLEtwo}
P=\lim_{y\rightarrow\infty,\, x\rightarrow 0
}\frac{\langle\Phi(\xi)\ptho(x)\ptho(y)\rangle}{\langle\ptho(x)\ptho(y)\rangle}\,.
\end{equation}
In the limit $y\to\infty$ the denominator is trivial, going as
$y^{-2h_3}$ and serving only to make the whole expression finite.
According to CFT \cite{BPZ,bigyellow}, the correlator of the $\phi_3$ operator
in the numerator satisfies a \em third\em-order equation of the
form
\begin{equation}\label{BPZ3}
\Big(
2{\rm Re}\,\Big[\frac{1}{(\xi-x)^2}\frac{\dd}{\dd\xi}\Big]
+2\mu{\rm Re}\,\Big[\frac{1}{\xi-x}\frac{\dd}{\dd\xi}\Big]\frac{\p}{\p x}
-\lambda\frac{\p^3}{\p x^3}\Big)P=0\,,
\end{equation}
where $\mu$ and
$\lambda$ are defined by the level 3 null state condition,
$$
(L_{-3}+\mu L_{-2}L_{-1}+\lambda L_{-1}^{3})|\ptho\rangle=0.
$$
In Appendix~\ref{l3n}, it is shown that
\begin{equation}
\mu=-\frac{2}{h_{3}}\,,\qquad \lambda=\frac{1}{h_{3}(1+h_{3})}\,,
\end{equation}
where $h_3$ is the conformal scaling dimension of $\ptho$. A
derivation of Eq.~(\ref{BPZ3}) is included in
Appendix~\ref{Lminus3}. As with the one curve case, the function
$P$ is expected to depend only on its angle from the imaginary
axis, or equivalently, the variable $t=(u-x)/v$, with $u$ and $v$
coming from $\xi=u+\tr{i}v$. This can be used to write the partial
differential equation as the following ordinary differential
equation in $t$
\begin{equation}\label{ode}
\lambda\frac{\dd^{3} P}{\dd t^{3}}-\frac{2\mu
t}{t^{2}+1}\frac{\dd^{2}P}{\dd
t^{2}}+\frac{(3-\mu)t^2-(1+\mu)}{(t^2+1)^{2}}\frac{\dd P}{\dd
t}=0\,.
\end{equation}
After some algebra this may be further rewritten in terms of
$Q(t)\equiv\dd P(t)/\dd t$ and the variable $s\equiv -t^2$ as
\begin{displaymath}
s(1-s)\frac{\dd^{2}Q}{\dd
s^2}+\frac{\lambda-(\lambda-2\mu)s}{2\lambda}\frac{\dd Q}{\dd s}
+\frac{(3-\mu)s+(1+\mu)}{4\lambda(1-s)}Q=0\,,
\end{displaymath}
which is of Riemann form, see Chapter $4$ of \cite{wandw}, with the following exponents:
\begin{itemize}
  \item as $s\approx0$, $Q\sim s^{0}, \,s^{1/2}$\,,
  \item as $s\approx1$, $Q\sim (s-1)^{1-8/\kappa}, \,(s-1)^{-8/\kappa}$\,,
  \item as $s\approx\infty$, $Q\sim s^{-4/\kappa}, \,s^{-(24-\kappa)/2\kappa}$\,.
\end{itemize}
The solutions to the ordinary differential equation therefore take
the form
\begin{equation}\label{sol}
Q(t)=A\frac{\,_{2}F_{1}(\frac{1}{2}+\frac{4}{\kappa},1-\frac{4}{\kappa};\frac{1}{2};-t^2)
+B\,t\,_{2}F_{1}(1+\frac{4}{\kappa},\frac{3}{2}-\frac{4}{\kappa};\frac{3}{2};-t^2)}{(1+t^2)^{\frac{8}{\kappa}-1}}\,.
\end{equation}
\subsection{Boundary conditions and solutions} Recall from
Section~\ref{DiffEqn2Curves} that there are three possibilities for the
position of a point in the upper half plane: it may be to the left
of both curves, between them or to the right of them. Each case
corresponds to different boundary conditions on $P$. Consider
firstly the case that the point lies to the left of both curves.
The boundary conditions for this case are
\begin{align*}
\lim_{t\to-\infty}\pl(t)&=1\,,\\
\lim_{t\to+\infty}\pl(t)&=O(t^{-(24/\kappa-2)})\,.
\end{align*}
The first condition comes from insisting that the probability of a
point on the negative real axis being to the left of both curves
is one. The second condition is best understood with reference to
Figure~\ref{h51} below,
%%%%%%%%%%%%%%%%%%%%%%%%%%%%%%%%%%%%%%%%%%%%%%%%%
\begin{figure}[ht]
\centering
\includegraphics[width=0.55\textwidth]{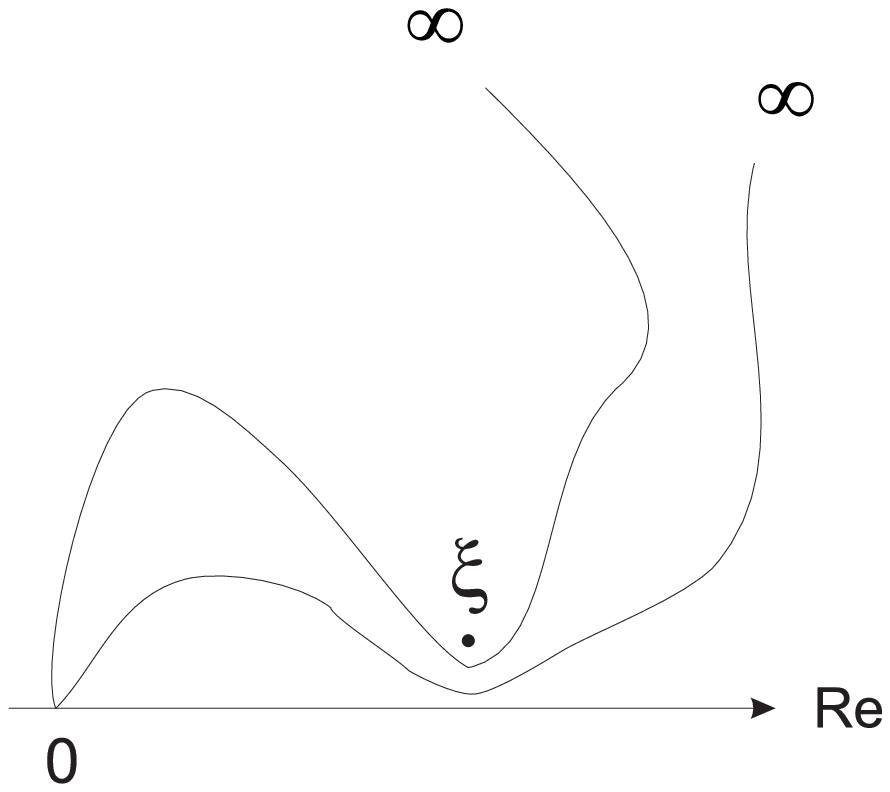}
\caption{$\pl$ as $t\rightarrow\infty$} \label{h51}
\end{figure}
%%%%%%%%%%%%%%%%%%%%%%%%%%%%%%%%%%%%%%%%%%%%%%%%%%
from which it can be seen that the limit $t\rightarrow\infty$
should have the exponent corresponding to a four-leg operator on
the real axis, namely $\tilde x_4=h_{5}=24/\kappa-2$.

Recall from the argument below Eq.~(\ref{ode}), that
$$
P=\int_{c}^{t}{Q(t')\dd t'}\,,
$$
with $Q(t)$ given in Eq.~(\ref{sol}) and $c$ a constant, to
be determined by the boundary conditions. In order to apply the
boundary conditions at large $t$, the hypergeometric functions
must be analytically continued for $|t|>1$, see for example
$15.3.8$ in~\cite{AbSt}. The expression for $P$ then takes the
form
\begin{align}
P(t)&=A\int_{c}^{t}\Big[\frac{\Gamma(\frac{1}{2})\Gamma(\frac{1}{2}-
\frac{8}{\kappa})}{\Gamma(1-\frac{4}{\kappa})\Gamma(-\frac{4}{\kappa})}
\frac{1}{(1+u^2)^{\frac{12}{\kappa}-\frac{1}{2}}}\,_{2}F_{1}(\frac{1}{2}+\frac{4}{\kappa},
\frac{4}{\kappa}-\frac{1}{2};\frac{8}{\kappa}+\frac{1}{2};\frac{1}{1+u^2})\nonumber\\
&+\frac{\Gamma(\frac{1}{2})\Gamma(\frac{8}{\kappa}-\frac{1}{2})}{\Gamma(\frac{1}{2}+
\frac{4}{\kappa})\Gamma(\frac{4}{\kappa}-\frac{1}{2})}\frac{1}{(1+u^2)^{\frac{4}{\kappa}}}
\,_{2}F_{1}(1-\frac{4}{\kappa},-\frac{4}{\kappa};-\frac{8}{\kappa}+\frac{3}{2};\frac{1}{1+u^2})\nonumber\\
&+B\frac{\Gamma(\frac{3}{2})\Gamma(\frac{1}{2}-\frac{8}{\kappa})}{\Gamma(\frac{3}{2}-
\frac{4}{\kappa})\Gamma(\frac{1}{2}-\frac{4}{\kappa})}\frac{u}{(1+u^2)^{\frac{12}{\kappa}}}
\,_{2}F_{1}(1+\frac{4}{\kappa},\frac{4}{\kappa};\frac{1}{2}+\frac{8}{\kappa};\frac{1}{1+u^2})\nonumber\\
&+B\frac{\Gamma(\frac{3}{2})\Gamma(\frac{8}{\kappa}-\frac{1}{2})}{\Gamma(1+\frac{4}{\kappa})
\Gamma(\frac{4}{\kappa})}\frac{u}{(1+u^2)^{\frac{4}{\kappa}+\frac{1}{2}}}\,_{2}F_{1}
(\frac{3}{2}-\frac{4}{\kappa},\frac{1}{2}-\frac{4}{\kappa};\frac{3}{2}-\frac{8}{\kappa};\frac{1}{1+u^2})\nonumber
\Big]\dd u\,.
\end{align}
The first and third terms in this large $t$ expansion approach
zero as $t^{2-24/\kappa}$, while the second and fourth term
approach zero as $t^{1-8/\kappa}$. For the range $0<\kappa<8$,
which is the range of physical interest, the first and third terms
fall off more quickly with increasing $t$. In order to satisfy
the second boundary condition, the coefficients of the dominant
second and fourth terms must ensure cancellation as $t\rightarrow\infty$.
This uniquely determines the constant $B$ as
$$
B=-2\frac{\Gamma(1+\frac{4}{\kappa})\Gamma(\frac{4}{\kappa})}{\Gamma
(\frac{1}{2}+\frac{4}{\kappa})\Gamma(-\frac{1}{2}+\frac{4}{\kappa})}\,.
$$
The solution follows immediately,
$$
\pl(t)=\frac{\int_{t}^{\infty}{S(t')\dd
t'}}{\int_{-\infty}^{\infty}{S(t')\dd t'}}\,,
$$
where
\begin{align*}
S(t)=\frac{\,_{2}F_{1}(\frac{1}{2}+\frac{4}{\kappa},1-\frac{4}{\kappa};\frac{1}{2};-t^2)
-\frac{2\Gamma(1+\frac{4}{\kappa})\Gamma(\frac{4}{\kappa})}{\Gamma
(\frac{1}{2}+\frac{4}{\kappa})\Gamma(-\frac{1}{2}+\frac{4}{\kappa})}\,
t\,_{2}F_{1}(1+\frac{4}{\kappa},\frac{3}{2}-\frac{4}{\kappa};\frac{3}{2};-t^2)}{(1+t^2)^{\frac{8}{\kappa}-1}}\,.
\end{align*}
It may be simplified using the identity
$$
\int_{-\infty}^{\infty}{S(t')\dd t'}=\frac{2^{2-8/\kappa}\pi
\Gamma(\frac{12}{\kappa}-1)}{\Gamma(\frac{4}{\kappa})\Gamma(\frac{8}{\kappa})}\,,
$$
see for example Eq.~($7.512.10$) in \cite{gradryz}. The solution then takes
the more elegant form
\begin{equation}\label{solgl}
\pl(t)=\frac{\Gamma(\frac{4}{\kappa})\Gamma(\frac{8}{\kappa})}
{2^{2-8/\kappa}\pi\Gamma(\frac{12}{\kappa}-1)}
\int_{t}^{\infty}S(t')\dd t'\,.
\end{equation}
It is a simple matter to derive $\pr$ since $\pr(t)=\pl(-t)$, thus
\begin{equation}\label{solgr}
\pr(t)=\frac{\Gamma(\frac{4}{\kappa})\Gamma(\frac{8}{\kappa})}
{2^{2-8/\kappa}\pi\Gamma(\frac{12}{\kappa}-1)}
\int_{-t}^{\infty}S(t')\dd t'\,.
\end{equation}
The remaining solution is $\pmm$. This can be derived in
two ways. Firstly, for the total probability to be unity,
$\pl+\pmm+\pr=1$. Subtracting the two previous solutions from one
leads to
\begin{equation}\label{solgm}
\pmm(t)=1-\frac{\Gamma(\frac{4}{\kappa})\Gamma(\frac{8}{\kappa})}
{2^{2-8/\kappa}\pi\Gamma(\frac{12}{\kappa}-1)}\Big[
\int_{t}^{\infty}S(t')dt'+\int_{-t}^{\infty}S(t')dt'\Big]\,.
\end{equation}
Alternatively, the solution can be written as the integral of the
odd term in $S(t)$ only, since we expect the solution to $\pmm$ to be an even
function of $t$. Applying the boundary condition
$\pmm(t\rightarrow\infty)=0$ determines $\pmm$ up to a
multiplicative constant
\begin{equation}\label{solgm2}
\pmm(t)=D\Big[1-\frac{\int_{0}^{t}\frac{u}{(1+u^2)^{\frac{8}{\kappa}-1}}\,_{2}F_{1}(1+\frac{4}{\kappa}
,\frac{3}{2}-\frac{4}{\kappa};\frac{3}{2};-u^2)du}{\int_{0}^{\infty}\frac{u}{(1+u^2)^{\frac{8}{\kappa}-1}}\,_{2}F_{1}(1+\frac{4}{\kappa}
,\frac{3}{2}-\frac{4}{\kappa};\frac{3}{2};-u^2)du}\Big]\,.
\end{equation}
The constant $D$ may be found by equating this expression with
Equation~(\ref{solgm}) at $t=0$
$$
D=1-\frac{\Gamma(\frac{4}{\kappa})\Gamma(\frac{8}{\kappa})}
{2^{1-8/\kappa}\pi\Gamma(\frac{12}{\kappa}-1)}\int_{0}^{\infty}S(t')dt'\,.
$$

\begin{figure}[ht]
\centering
\includegraphics[width=0.95\textwidth]{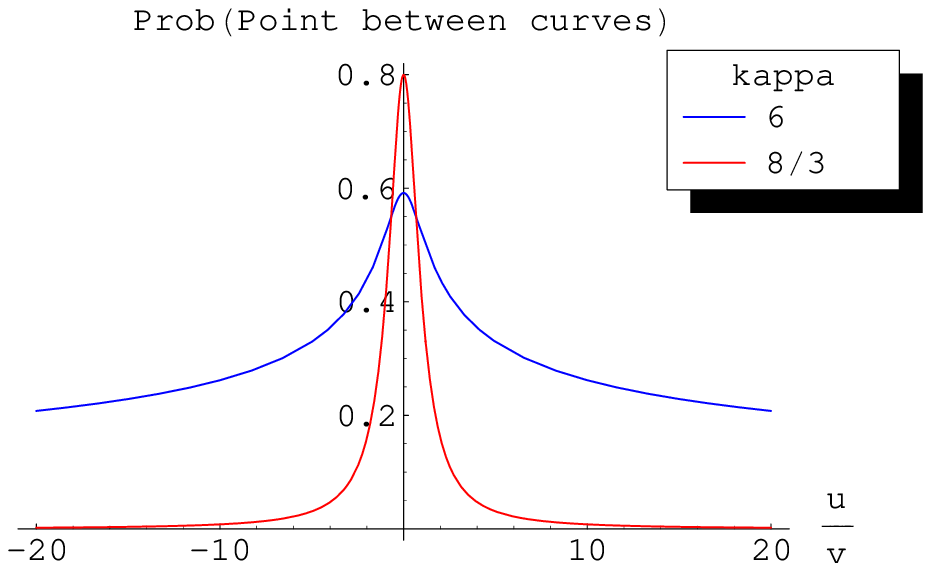}
\includegraphics[width=0.95\textwidth]{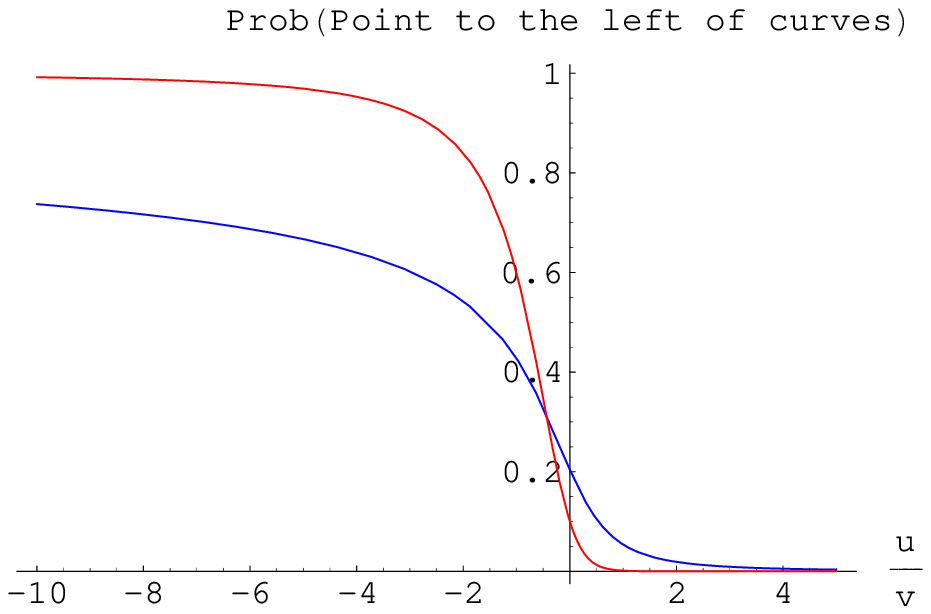}
\caption{The self-avoiding walk ($\kappa=8/3$) and percolation
($\kappa=6$)} \label{midleft}
\end{figure}

\newpage
Below, in Figure~(\ref{midleft}) are the results for $\kappa=8/3$,
conjecturally corresponding to the scaling limit of two mutually
avoiding self-avoiding walks starting near the origin, and
$\kappa=6$, corresponding to the scaling limit of the percolation
problem.
%%%%%%%%%%%%%%%%%%%%%%%%%%%%%%%%%%%%%%%%%%%%%%%%%
%%%%%%%%%%%%%%%%%%%%%%%%%%%%%%%%%%%%%%%%%%%%%%%%%%

\newpage
\section{Special Cases}\label{results}
In the special cases $\kappa=0,2,8/3,4,8$, explicit solutions may
be derived for $\pl$, $\pmm$ and $\pr$.

\subsection{$\kappa=0$} \label{kappa0} 
\begin{figure}[h]
\centering
\includegraphics[width=0.55\textwidth]{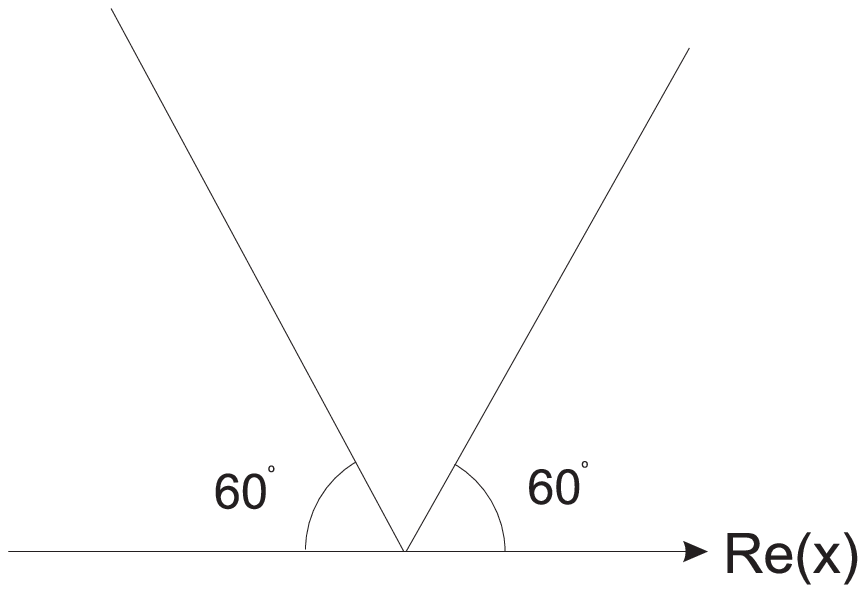}
\caption{$\kappa=0$} \label{kzero}
\end{figure}

With $\kappa=0$, the
curves are deterministic straight lines. They start at the origin
and proceed at an angle of $\pi/3$ radians from each other and
from the real axis, as in Figure~(\ref{kzero}). Actually this is a
rather singular limit of the equations, and the above result is
more easily understood following the multiple SLE interpretation
of Sec.~\ref{RevEng}.
%%%%%%%%%%%%%%%%%%%%%%%%%%%%%%%%%%%%%%%%%%%%%%%%%
%%%%%%%%%%%%%%%%%%%%%%%%%%%%%%%%%%%%%%%%%%%%%%%%%%
\subsection{$\kappa=2$} The solutions are written in terms of the
function $S(t)$. In the case of $\kappa=2$, this takes the form
\begin{align*}
S(t)&=\frac{1}{(1+t^2)^3}\Big[
\,_{2}F_{1}(\frac{5}{2},-1,\frac{1}{2},-t^2)
-\frac{32}{3\pi}t\,_{2}F_{1}(3,-\frac{1}{2},\frac{3}{2},-t^2)\Big]\\
&=\frac{1}{(1+t^2)^3}\Big[1+5t^2-\frac{32
t}{3\pi}\Big(\frac{t(13+15
t^2)+3(1+6t^2+5t^4)\atan(t)}{16(t+t^3)}\Big) \Big]\,.
\end{align*}
Inserting this expression into
Eq.~(\ref{solgl}),~(\ref{solgr})~and~(\ref{solgm}) and
performing the integrations, the following solutions are obtained
\begin{align}
\pl(t)&=\frac{1}{4}+\frac{1}{9\pi^2(1+t^2)^3}\Big[
(-16-9t^2+9t^4)-9\pi(t^3+t^5)\nonumber\\
&\quad+9(1+t^2)\atan(t)\Big(2t^3-\pi(1+t^2)^2+(1+t^2)^2\atan(t)\Big)\Big]
\\
\pr(t)&=\frac{1}{4}+\frac{1}{9\pi^2(1+t^2)^3}\Big[
(-16-9t^2+9t^4)+9\pi(t^3+t^5)\nonumber\\
&\quad+9(1+t^2)\atan(t)\Big(2t^3+\pi(1+t^2)^2+(1+t^2)^2\atan(t)\Big)\Big]
\\
\pmm(t)&=\frac{1}{2}-\frac{2}{9\pi^2(1+t^2)^3}\Big[
(-16-9t^2+9t^4)\nonumber\\
&\quad+9(1+t^2)\atan(t)\Big(2t^3+(1+t^2)^2\atan(t)\Big)\Big]\,.
\end{align}

\subsection{$\kappa=8/3$} Analytic solutions may be obtained for
the special case of the scaling limit of two self-avoiding walks,
conjectured to correspond to $\kappa=8/3$. In this case, the
function $S(t)$ is given by
\begin{align*}
S(t)&=\frac{1}{(1+t^2)^2}\Big[\,_{2}F_{1}(2,-\frac{1}{2};\frac{1}{2};-t^2)-\frac{3\pi
t}{4} \Big]\\
&=\frac{1}{(1+t^2)^2}\Big(
-\frac{1}{2(1+t^2)}+\frac{3t}{2}\arctan(t)+\frac{3}{2}-\frac{3\pi t}{4}
\Big)
\,.
\end{align*}
The solution for $\pl$ follows as
\begin{align*}
\pl(t)&=\frac{\Gamma(\frac{4}{\kappa})\Gamma(\frac{8}{\kappa})}
{2^{2-8/\kappa}\pi\Gamma(\frac{12}{\kappa}-1)}
\int_{t}^{\infty}S(t')\dd t'\\
&=\frac{16}{15\pi}\int_{t'}^{\infty}{\frac{1}{(1+t'^2)^2}\Big(
-\frac{1}{2(1+t'^2)}+\frac{3t'}{2}\arctan(t')+\frac{3}{2}-\frac{3\pi t'}{4}
\Big)\dd t'}\\
&=\frac{16}{15\pi}\Big[
\frac{t'(13+15t'^2)+3(1+6t'^2+5t'^{4})\arctan(t')}{16(1+t'^2)^2}+\frac{3\pi}{8}\frac{1}{1+t'^2}
\Big]^{\infty}_{t}\\
&=\frac{1}{2}-\frac{16}{15\pi}\frac{t(13+15t^2)+3(1+6t^2+5t^4)\atan(t)}
{16(1+t^2)^2}-\frac{2}{5}\frac{1}{1+t^2}\,.
\end{align*}
This may be simplified to
\begin{equation}
\pl(t)=\frac{
-2t(13+15t^2)+3\pi(1+6t^2+5t^4)-6(1+6t^2+5t^4)\atan(t)}{30\pi(1+t^2)^{2}}\,.
\end{equation}
The other solutions are
\begin{align}
&\pr(t)=\frac{
2t(13+15t^2)+3\pi(1+6t^2+5t^4)+6(1+6t^2+5t^4)\atan(t)}{30\pi(1+t^2)^{2}}\\
&\pmm(t)=\frac{4}{5(1+t^2)}\,.
\end{align}

\subsection{$\kappa=4$} With $\kappa=4$, $S(t)$ takes the form
$$
S(t)=\frac{1}{1+t^2}\Big[
\,_{2}F_{1}(\frac{3}{2},0,\frac{1}{2},-t^2)+Bt\,_{2}F_{1}(2,\frac{1}{2},\frac{3}{2},-t^2)\Big]\,,
$$
where $B=-4/\pi$. Substituting for the hypergeometric functions in
terms of elementary functions leads to
$$
S(t)=\frac{1}{1+t^2}\Big[
1-\frac{2t}{\pi}\Big(\frac{1}{1+t^2}+\frac{\atan(t)}{t}\Big)
\Big]\,.
$$
Inserting this form into the expression for $\pl$
\begin{equation}
\pl(t)=\frac{1}{4}-\frac{1}{\pi^2(1+t^2)}-\frac{\atan(t)}{\pi}+\frac{\atan^{2}(t)}{\pi^2}\,.
\end{equation}
The other solutions are
\begin{align}
\pr(t)&=\frac{1}{4}-\frac{1}{\pi^2(1+t^2)}+\frac{\atan(t)}{\pi}+\frac{\atan^{2}(t)}{\pi^2}\\
\pmm(t)&=\frac{1}{2}+\frac{2}{\pi^2(1+t^2)}-\frac{2\atan^{2}(t)}{\pi^2}\,.
\end{align}
In terms of the angle $\phi$ from the imaginary axis to the point
$\xi$, which is to say $\phi=\atan(u/v)=\atan(t)$,
\begin{align*}
\pl(\phi)&=\frac{1}{4} -\frac{\textrm{cos}^{2}(\phi)}{\pi^2}
-\frac{\phi}{\pi} +\frac{\phi^2}{\pi^2}\\
\pr(\phi)&=\frac{1}{4} -\frac{\textrm{cos}^{2}(\phi)}{\pi^2}
+\frac{\phi}{\pi}
+\frac{\phi^2}{\pi^2}\\
\pmm(\phi)&=\frac{1}{2}+\frac{2\cos^2(\phi)}{\pi^2}-\frac{2\phi^2}{\pi^2}\,.
\end{align*}

\subsection{$\kappa=8$}

The case $\kappa=8$ is subtle and requires careful treatment. For the single curve,
taking the limit as $\kappa\to8-$ at fixed $t$
yields an expression which
fails to satisfy the boundary conditions \cite{Schramm}. Instead, the
probability of any point not on the real axis
being to one side of the curve is everywhere equal to a half.
This however makes physical sense since for $\kappa=8$ the curve is
space-filling.

For the case of two curves
there is a similar boundary condition violating solution
which is obtained by continuing the general solution for finite $t$
to $\kappa=8$.
There is a second solution, however, which follows from solving the differential equations at
$\kappa=8$ and which does satisfy the boundary conditions.

First, consider the general solution in the limit $\kappa\rightarrow 8$. Using the usual definition
\begin{equation*}
S(t)=\frac{\,_{2}F_{1}(\frac{1}{2}+\frac{4}{\kappa},1-\frac{4}{\kappa};\frac{1}{2};-t^2)
-\frac{2\Gamma(1+\frac{4}{\kappa})\Gamma(\frac{4}{\kappa})}{\Gamma
(\frac{1}{2}+\frac{4}{\kappa})\Gamma(-\frac{1}{2}+\frac{4}{\kappa})}\,
t\,_{2}F_{1}(1+\frac{4}{\kappa},\frac{3}{2}-\frac{4}{\kappa};\frac{3}{2};-t^2)}{(1+t^2)^{\frac{8}{\kappa}-1}}\,,
\end{equation*}
the solution is
\begin{align*}
\pl(t)&=\frac{\int_{t}^{\infty}{S(t')\dd
t'}}{\int_{-\infty}^{\infty}{S(t')\dd t'}}\\
&=\frac{\Gamma(\frac{4}{\kappa})\Gamma(\frac{8}{\kappa})}
{2^{2-8/\kappa}\pi\Gamma(\frac{12}{\kappa}-1)}
\int_{t}^{\infty}S(t')\dd t'\,.
\end{align*}
In anticipation of taking the limit $\kappa\rightarrow 8$, define $\kappa=8-\epsilon$ and
assume that $\epsilon$ is small. Rewriting the above expressions in terms
of $\epsilon$:
\begin{align*}
&S(t)=\frac{\,_{2}F_{1}(1+\es,\frac{1}{2}-\es;\frac{1}{2};-t^2)
-\frac{2\g(\frac{3}{2}+\es)\g(\frac{1}{2}+\es)}{\g(1+\es)\g(\es)}\,
t\,_{2}F_{1}(\frac{3}{2}+\es,1-\es;\frac{3}{2};-t^2)}
{(1+t^2)^{\frac{\epsilon}{8}}}\\
&\pl(t)=\frac{\g(\h+\es)\g(1+\ee)}{2^{1-\ee}\pi\g(\h+\tes)}
\int_{t}^{\infty}S(t')\dd t'\,.
\end{align*}
The two hypergeometric functions in $S(t)$ can be analytically continued for $t>0$
\begin{align*}
&S(t)=\frac{\sqrt{\pi}\g(-\frac{1}{2}-\ee)}{\g(\h-\es)\g(-\h-\es)}
\frac{1}{(1+t'^2)^{1+\tes}}\,_{2}F_{1}(1+\es,\es;\frac{3}{2}+\ee;\frac{1}{1+t'^2})\\
&+\frac{\sqrt{\pi}\g(\h+\ee)}{\g(1+\es)\g(\es)}\frac{1}{(1+t'^2)^{\h+\es}}\,_{2}F_{1}(\h-\es,-\h-\es;\h-\ee;\frac{1}{1+t'^2})\\
&-\sqrt{\pi}\frac{\g(\thh+\es)\g(\h+\es)\g(-\h-\ee)}{\g(1+\es)\g(\es)\g(1-\es)\g(-\es)}
\frac{t'\,_{2}F_{1}(\thh+\es,\h+\es;\thh+\ee;\frac{1}{1+t'^2})}{(1+t'^2)^{\thh+\tes}}\\
&-\sqrt{\pi}\frac{\g(\h+\ee)}{\g(1+\es)\g(\es)}\frac{t'\,_{2}F_{1}(1-\es,-\es;\h-\ee;\frac{1}{1+t'^2})}{(1+t'^2)^{1+\es}}
\,.
\end{align*}
Each of these four terms are to be integrated from $t'=t$ to $\infty$. This can be done by expanding the
hypergeometric functions and integrating term by term. Consider the integral of the first function in $S(t)$ for $t>0$
\begin{align*}
&\frac{\sqrt{\pi}\g(-\frac{1}{2}-\ee)}{\g(\h-\es)\g(-\h-\es)}
\int_{t}^{\infty}{\frac{1}{(1+t'^2)^{1+\tes}}\,_{2}F_{1}(1+\es,\es;\frac{3}{2}+\ee;\frac{1}{1+t'^2})}\dd t'\\
&=\frac{\sqrt{\pi}\g(-\frac{1}{2}-\ee)}{\g(\h-\es)\g(-\h-\es)}
\int_{t}^{\infty}{\frac{1}{(1+t'^2)^{1+\tes}}\Big[1+\frac{(1+\es)(\es)}{(\thh+\ee)1!}\frac{1}{1+t^2}+\ldots\Big]}\dd t'\\
&=\frac{\sqrt{\pi}\g(-\frac{1}{2}-\ee)}{\g(\h-\es)\g(-\h-\es)}
\int_{t}^{\infty}{\frac{1}{(1+t'^2)^{1+\tes}}\dd t'}+\epsilon f(\epsilon,t)\\
&=-t\,_{2}F_{1}(\h,1+\tes;\thh;-t^2)+\frac{\sqrt{\pi}\g(\h+\tes)}{2\g(1+\tes)}
+\epsilon f(t,\epsilon)
\,,
\end{align*}
where $f(\epsilon,t)$ is a convergent function of its arguments for all $t$ in the limit $\epsilon\rightarrow 0$.

After integration, the third term gives $\epsilon^2 g(\epsilon,t)$, where
$g(\epsilon,t)$ is also a convergent function of its arguments for all $t$ in the limit $\epsilon\rightarrow 0$.

The integral of the second and fourth terms should be considered together. They contribute
\begin{align*}
\frac{\sqrt{\pi}\g(\h+\ee)}{\g(1+\es)\g(\es)}\int_{t}^{\infty}{\Big[
\frac{1}{(1+t'^2)^{\h+\es}}-\frac{t'}{(1+t'^2)^{1+\es}}\Big]\dd t'}+\epsilon h(\epsilon,t)\,,
\end{align*}
where $h(\epsilon,t)$ is a convergent function for all $t$ in the limit $\epsilon\rightarrow 0$.
The integral of $S(t)$ is therefore
\begin{align*}
&\int_{t}^{\infty}{S(t')\dd t'}=-t\,_{2}F_{1}(\h,1+\tes;\thh;-t^2)+\frac{\sqrt{\pi}\g(\h+\tes)}{2\g(1+\tes)}+\epsilon f(t,\epsilon)+\epsilon^{2} g(t,\epsilon)\\
&\qquad\qquad+\frac{\sqrt{\pi}\g(\h+\ee)}{\g(1+\es)\g(\es)}\int_{t}^{\infty}{\Big[
\frac{1}{(1+t'^2)^{\h+\es}}-\frac{t'}{(1+t'^2)^{1+\es}}\Big]\dd t'}+\epsilon h(\epsilon,t)
\,.
\end{align*}
Taking the limit $\epsilon\rightarrow 0$, the first two terms become
$$
-\arctan(t)+\frac{\pi}{2}
$$
and the terms involving $f(\epsilon,t),g(\epsilon,t),h(\epsilon,t)$ have coefficients which tend to the zero.
The remaining term is
\begin{align*}
&\lim_{\epsilon\rightarrow 0}{\frac{\sqrt{\pi}\g(\h+\ee)}{\g(1+\es)\g(\es)}\int_{t}^{\infty}{\Big[
\frac{1}{(1+t'^2)^{\h+\es}}-\frac{t'}{(1+t'^2)^{1+\es}}\Big]\dd t'}}\,,
\end{align*}
where $j(\epsilon,t)$ is a convergent function of its arguments for all $t$ in the limit $\epsilon\rightarrow 0$.
The integral is finite, so the expression goes to zero as $\epsilon\rightarrow 0$.

Using
$$
\lim_{\epsilon\rightarrow 0}{\frac{\g(\h+\es)\g(1+\ee)}{2^{1-\ee}\pi\g(\h+\tes)}=\frac{1}{2\pi}}\,,
$$
allows the solution for $\pl(t)$ and related expressions to be
deduced in terms of the angle $\phi$ defined in the previous
subsection,
\begin{align}
\pl(\phi)&=\frac{1}{4}-\frac{\phi}{2\pi}\,,\\
\pr(\phi)&=\frac{1}{4}+\frac{\phi}{2\pi}\,,\\
\pmm(\phi)&=\frac{1}{2}\,.
\end{align}
Note that these are valid only for $\phi\not=\pm\pi/2$. The
limits $\kappa\to8$ and $|t|\to\infty$ do not commute.

Now take $\kappa=8$ from the beginning. The differential equation has solutions
of the form
$$
\pl(t)=A\arctan(t)+B\,.
$$
If $A$ and $B$ are chosen to be $-1/\pi$ and $1/2$ respectively, this satisfies the boundary conditions.
Then in terms of the angle $\phi$,
\begin{align}
\pl(\phi)&=\frac{1}{2}-\frac{\phi}{\pi}\,,\\
\pr(\phi)&=\frac{1}{2}+\frac{\phi}{\pi}\,,\\
\pmm(\phi)&=0\,.
\end{align}
This is not the analytic continuation of the first solution to $\kappa=8$.
Mathematically, this may be traced to the fact that the limits
$\kappa\to8-$ and $t\to +\infty$ do not commute, yet we have to impose the
boundary condition on $\pl$ at $t= +\infty$.

Physically, these two solutions appear to lead to different pictures.
In the first case, the probability that a given point lies between the
two curves is exactly $\frac12$.
This may be understood\footnote{We are grateful to W.~Werner for
pointing this
out.} in terms of the physical
picture of the curves being the boundaries of two disjoint uniform
spanning clusters which are separated by some random simple curve
$\gamma_{\rm sep}$.
Each curve fills the entire region to the left(right) of
$\gamma_{\rm sep}$. Within each
region, the probability that a given point lies to the left(right)
of the curve is $\frac12$ as before, but the probability that it
lies in this region, to the left(right) of the separatrix is
$\frac12(1\mp2\phi/\pi)$. Thus the probability that
a point lies to the left of both curves is $\h\cdot\h(1-2\phi/\pi)$, while
the probability this it lies between them is
\begin{equation}
\h\cdot\h(1-2\phi/\pi)+\h\cdot\h(1+2\phi/\pi)=\h\,.
\end{equation}

In the second solution, the probability that a point is between
the curves is zero, hence the area between them vanishes.
Physically, this could correspond to two mutually avoiding dense
polymers, which also correspond to $\kappa=8$. In this case,
however, they find it entropically favourable closely to follow
each other. The probability that a point lies to one side of this
composite curve is the same as for a single curve with $\kappa=4$,
see \cite{Schramm}. This may be understood in terms of the
stochastic interpretation of Sec.~\ref{RevEng}. For $\kappa=8$ the
points $x_1$ and $x_2$ almost certainly collide in finite time,
after which it is necessary to prescribe how to continue the
solution. One possibility is that the two points coalesce, in
which case their centre of mass describes a Brownian motion with
$\kappa'=\frac12\kappa=4$, as in the second solution. Another
possibility is that they are conditioned never to collide, which
then presumably corresponds to the first case.

\section{Application to the quantum Hall transition}\label{qhall}
The Quantum Hall effect is observed in two dimensional electron
gases in semiconductors with magnetic fields applied normal to the
plane. Donor ions are spatially separated from the electron gas in
order to increase the mobility of the electrons in the sample. The
ions are positively charged and the resulting Coulomb potential in
the electron gas can be modelled as a random potential, $V$. A
semi-classical approximation may be used in the limits of slowly
varying potential on the scale of the magnetic length
$l=\sqrt{\hbar c/eB}$ and strong magnetic field, defined as
$$
\hbar \omega_{c}\equiv\frac{\hbar eB}{mc}\ll \epsilon_{F}\,,
$$
where $\epsilon_{F}$ is the Fermi energy of the system. In this
limit, the eigenfunctions of electrons are large only around
constant energy surfaces of the potential, $V$, see
\cite{Trugman}. The eigenfunctions can be approximated by
$$
\psi(u,v)=C(u)\chi_{n}(v)e^{i\phi(u,v)}\,,
$$
where $u$ is the length along the constant energy surface and $v$
is the distance normal to it. $C$ is a normalisation factor and
$\chi_{n}$ is the $n$th harmonic oscillator function
$$
\chi_{n}=H_{n}(\frac{v}{l})e^{-\frac{v^2}{2l^2}}\,.
$$

Choose the zero of the random potential to be the spatial average
of the potential and assume that $\langle V(r)V(r+\delta r)\rangle$ goes to
zero for $\delta r\gg b$. The requirement that the potential varies slowly
compared to the magnetic length is equivalent to $b>>l$.
If all points where the potential is greater than a
value $E$ are coloured white and points where $V\leq E$ are
coloured black, the lines of constant potential will be given by
the boundary between the two coloured regions. This model is
believed to be in the same universality class as lattice
percolation. Thus, electrons will move along the boundaries of
percolation clusters. In general, the boundaries will form closed
loops, but for the critical value $E=0$, their mean size diverges
and they will be locally described by SLE with $\kappa=6$.
The appropriate boundary conditions for the percolation picture are
that the top and bottom edges should be coloured white since the
potential is effectively infinite there.

%%%%%%%%%%%%%%%%%%%%%%%%%%%%%%%%%%%%%%%%%%%%%%%%%
\begin{figure}[ht]
\centering
\includegraphics[width=0.9\textwidth]{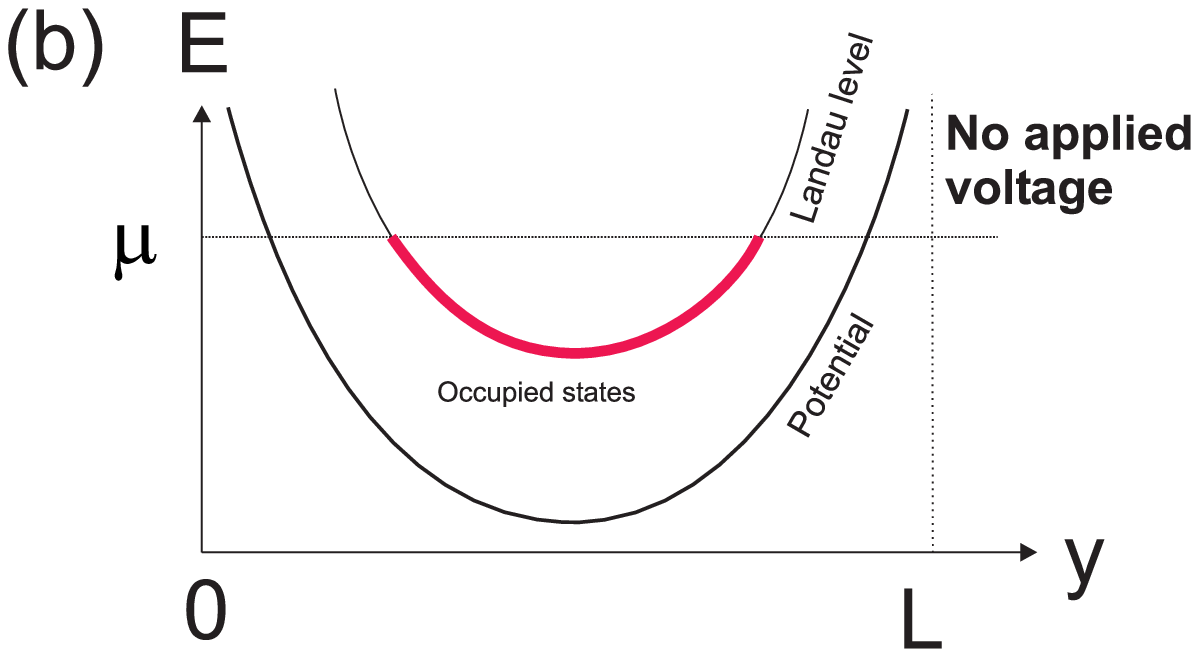}
\includegraphics[width=0.9\textwidth]{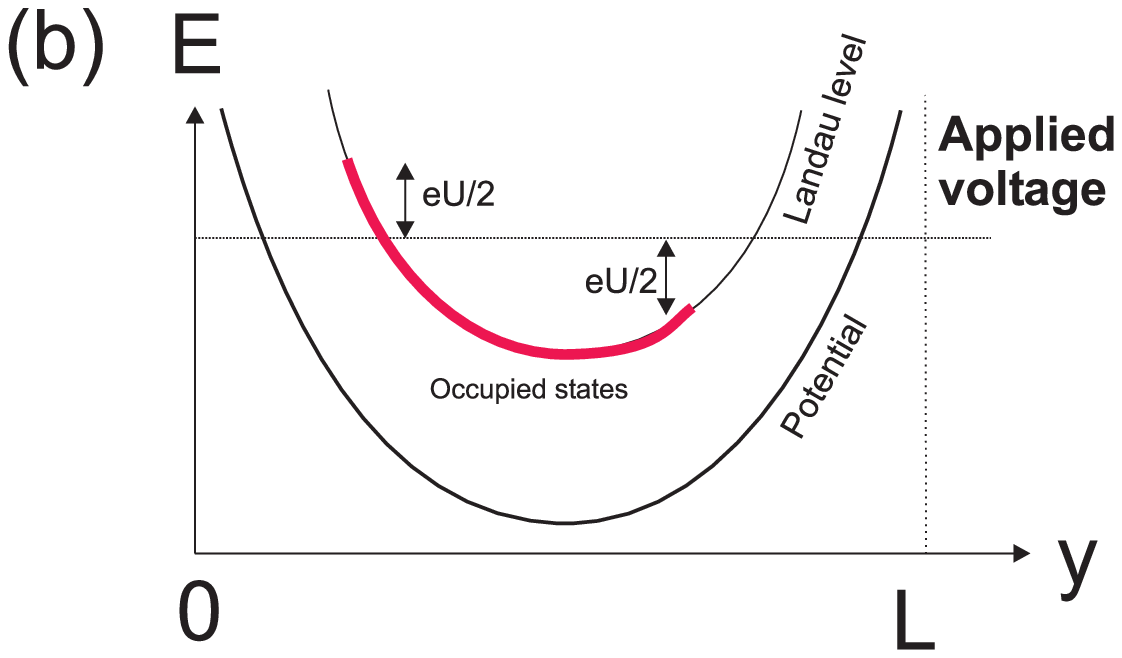}
\caption{The spatial distribution of eigenstates, showing occupation before and
after a potential difference is applied between the ends of the sample.} \label{potential}
\end{figure}
%%%%%%%%%%%%%%%%%%%%%%%%%%%%%%%%%%%%%%%%%%%%%%%%%%

\begin{figure}[ht]
\centering
\includegraphics[width=0.9\textwidth]{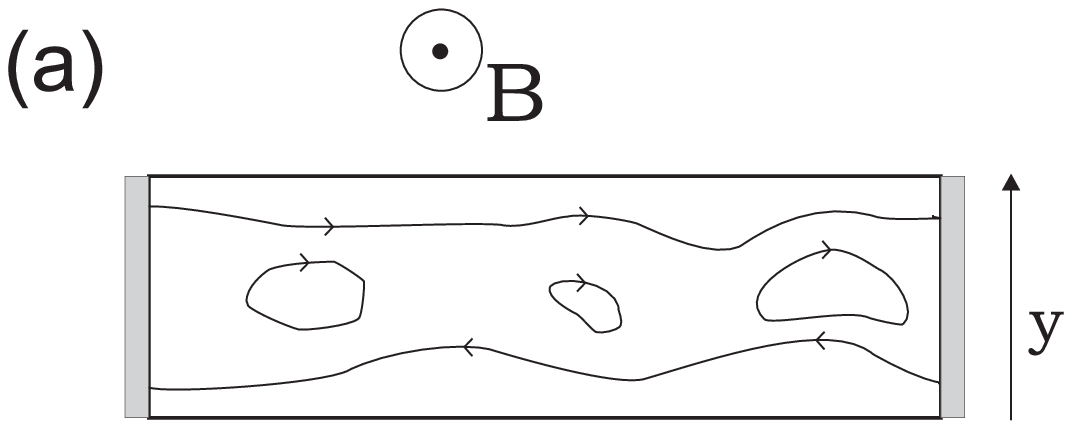}
\includegraphics[width=0.9\textwidth]{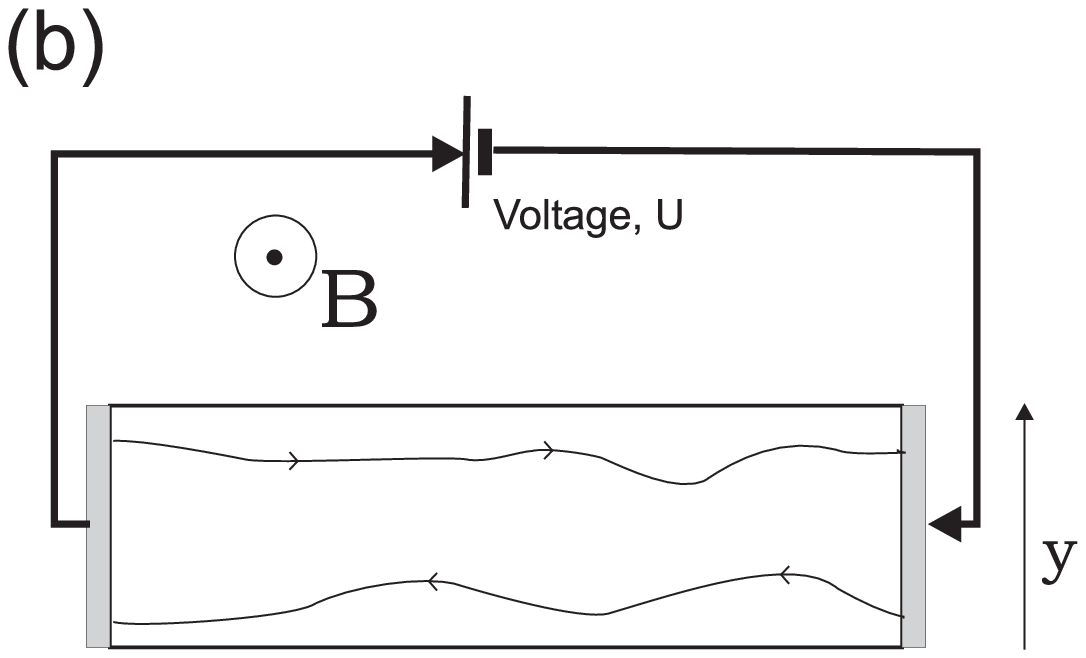}
\caption{The top diagram depicts the diamagnetic currents. The figure below shows the
Hall bar with leads attached and a voltage applied. The arrows in both diagrams
represent the direction of electron flow.} \label{geom}
\end{figure}

\begin{figure}[h]
\centering
\includegraphics[width=0.75\textwidth]{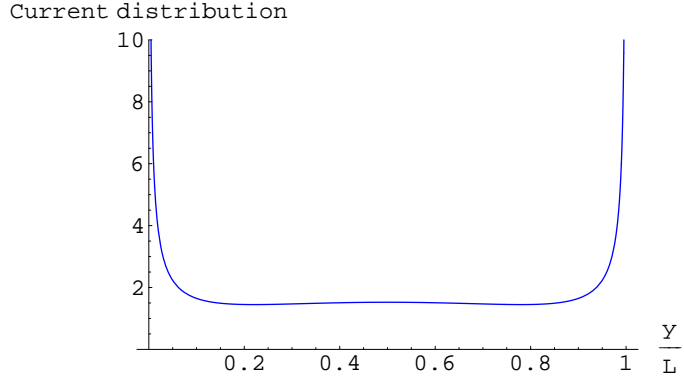}
\caption{Average current distribution for the continuum
percolation model} \label{currentdistr}
\end{figure}

As a function of $y$, the distance across the sample, the energy of the Landau levels follow
the form of the potential and those states with $E\leq\mu$, the chemical potential,
are occupied, as shown in Figure~(\ref{potential}a).
Diamagnetic currents flow, both around the closed loops and along
the extended cluster boundaries with $E=0$, as shown in Figure~(\ref{geom}a).
On connection of the leads to the ends and the application of a
potential difference, the current
distribution will change from that of the equilibrium case.
%%%%%%%%%%%%%%%%%%%%%%%%%%%%%%%%%%%%%%%%%%%%%%%%%
%%%%%%%%%%%%%%%%%%%%%%%%%%%%%%%%%%%%%%%%%%%%%%%%%%
The currents flowing along the extended cluster boundaries will be
affected the most, since these are the only paths which can carry
a net current along the length of the sample without potential
tunnelling. The Fermi energy of electrons moving along the bottom
cluster will be the equilibrium level plus $eU/2$, since these
electrons flow from the electron reservoir at the negative battery
contact. The Fermi energy of the electrons in the upper cluster
boundary will be that of the equilibrium case minus $eU/2$, since
these electrons flow from the electron reservoir at the positive
battery contact. Relative to the equilibrium case, therefore, an
extra current, $\delta I$ flows along the lower cluster boundary
and, to a first approximation, $\delta I$ less current flows along
the upper cluster boundary. The averaged change in the current
distribution, compared to the equilibrium case, will be given by
the spatial average of the extended cluster boundaries in the
percolation picture ($\kappa=6$).
%%%%%%%%%%%%%%%%%%%%%%%%%%%%%%%%%%%%%%%%%%%%%%%%%
%\begin{figure}[h]
%\centering
%\includegraphics[width=0.55\textwidth]{qperc.eps}
%\caption{A percolation picture of the QHE} \label{qperc}
%\end{figure}
%%%%%%%%%%%%%%%%%%%%%%%%%%%%%%%%%%%%%%%%%%%%%%%%%%
We may relate the strip geometry of the Hall bar experiment to the half-plane
discussed in this paper by the conformal mapping $z\rightarrow z'=w(z)$ with
$$
w(z)=\frac{L}{\pi}\ln(z)\,,
$$
which maps the upper half plane to an infinite strip of width $L$.
Lines with constant angle, parametrised by their value of $t=u/v$,
are mapped to lines of constant distance, $y$, from the bottom of
the infinite strip, given by
$$
y=\frac{L}{\pi}\arctan(\frac{1}{t})\,.
$$
A little
thought then shows that the mean extra current density flowing along the upper
boundary curve at height $y$ in
the sample is proportional to the derivative with respect to $y$
of the probability that $y$ is above both curves. Similarly, the extra
current density flowing along the lower curve is $-\dd P_{\textrm{below}}/\dd y$.
Thus,
\begin{align*}
I&\propto
\frac{\dd \pl}{\dd
y}-\frac{\dd \pr}{\dd
y}\\&\propto-\frac{\Gamma(\frac{4}{\kappa})\Gamma(\frac{8}{\kappa})}
{2^{2-8/\kappa}\pi\Gamma(\frac{12}{\kappa}-1)}\frac{\pi}{L}
(1+\frac{1}{\tan(\frac{\pi y}{L})^2})\Big[S(\frac{1}{\tan(\frac{\pi
y}{L})})+S(\frac{1}{\tan(\pi-\frac{\pi
y}{L})})\Big]\,,
\end{align*}

where, as previously,
\begin{align*}
S(t)=\frac{\,_{2}F_{1}(\frac{1}{2}+\frac{4}{\kappa},1-\frac{4}{\kappa};\frac{1}{2};-t^2)
-\frac{2\Gamma(1+\frac{4}{\kappa})\Gamma(\frac{4}{\kappa})}{\Gamma
(\frac{1}{2}+\frac{4}{\kappa})\Gamma(-\frac{1}{2}+\frac{4}{\kappa})}\,
t\,_{2}F_{1}(1+\frac{4}{\kappa},\frac{3}{2}-\frac{4}{\kappa};\frac{3}{2};-t^2)}
{(1+t^2)^{\frac{8}{\kappa}-1}}\,.
\end{align*}
For $\kappa=6$ this becomes
$$
I\propto\frac{\Gamma(\frac{2}{3})\Gamma(\frac{4}{3})}
{2^{2/3}L}(1+\frac{1}{\tan(\frac{\pi y}{L})^2})\Big[S(\frac{1}{\tan(\frac{\pi
y}{L})})+S(\frac{1}{\tan(\pi-\frac{\pi
y}{L})})\Big]\,,
$$
with
\begin{align*}
S(t)=\frac{\,_{2}F_{1}(\frac{5}{6},\frac{2}{3};\frac{1}{2};-t^2)
-\frac{2\Gamma(\frac{5}{3})\Gamma(\frac{2}{3})}{\Gamma
(\frac{7}{6})\Gamma(\frac{1}{6})}\,
t\,_{2}F_{1}(\frac{5}{3},\frac{5}{6};\frac{3}{2};-t^2)}{(1+t^2)^{\frac{1}{3}}}\,.
\end{align*}
A plot of the mean current distribution is presented in Figure
(\ref{currentdistr}).
%%%%%%%%%%%%%%%%%%%%%%%%%%%%%%%%%%%%%%%%%%%%%%%%%
%%%%%%%%%%%%%%%%%%%%%%%%%%%%%%%%%%%%%%%%%%%%%%%%%%
There are two important potential limitations on the applicability
of this result: (i) we have ignored Coulomb interactions between
electrons, which, although they are often assumed not to change
the single-electron picture of the plateaux transition, may well
affect the current distribution; (ii) we have neglected quantum
tunnelling between neighbouring regions of zero potential, which
are believed to be relevant and to change the universality class
away from classical percolation. However, for the spin quantum
Hall transition\cite{Gruzberg,CardyChalker} this is not the case.

\section{Reverse Engineering SLE$(\kappa,2)$}\label{RevEng}
Now let us consider the more general case of two curves starting at the points $x_{1}\neq x_{2}$.
$\pl$, $\pmm$ and $\pr$ are given by the ratio of correlation functions
\begin{align*}
P(x_{1},x_{2},\xi)&=\frac{\langle\Phi(\xi)\pto(x_{1})\pto(x_{2})\pto(\infty)\pto(\infty)\rangle}
{\langle\pto(x_{1})\pto(x_{2})\pto(\infty)\pto(\infty)\rangle}
\equiv\frac{F_{\Phi}(x_{1},x_{2},\xi)}{F_{\textbf{1}}(x_{1},x_{2})}\,,
\end{align*}
Using the following definition of the differential operator $D_{1}$:
$$
D_{1}\equiv\Big[ \frac{\kappa}{4}\frac{\p^2}{\p
x_{1}^2}-\frac{h_{2,1}}{(x_{2}-x_{1})^2}+\frac{1}{x_{2}-x_{1}}\frac{\p}{\p
x_{2}} \Big]\,,
$$
$F_{\Phi}$ satisfies the BPZ differential equation
\begin{align*}
D_{1}F_{\Phi}(x_{1},x_{2},\Phi)&=<\langle\delta_{x_{1}}\Phi(\xi)\pto(x_{1})\pto(x_{2})
\pto^2(\infty)\rangle\,,\\
&\textrm{with}\quad\delta_{x_{1}}=\frac{1}{\xi-x_{1}}
\,.
\end{align*}
Writing $F_{\Phi}=P(x_{1},x_{2},\Phi) F_{\textbf{1}}(x_{1},x_{2})$
and using $D_{1}F_{\textbf{1}}(x_{1},x_{2})=0$\,,
\begin{equation}\label{reven}
D_{1}P(x_{1},x_{2},\Phi)+\frac{\kappa}{2F_{\textbf{1}}(x_{1},x_{2})}\frac{\p
F_{\textbf{1}}(x_{1},x_{2})}{\p x_{1}}\frac{\p
P(x_{1},x_{2},\Phi)}{\p x_{1}} =\delta_{x_{1}}P(x_{1},x_{2},\Phi)\,.
\end{equation}
$F_{\textbf{1}}(x_{1},x_{2})$ is a three point function. This can
be seen by taking the fusion of the operators at infinity, ie
$\pto^{2}(\infty)=\ptho(\infty)$. Conformal field theory may be
employed to fix the form of the three point function up to a
constant
\begin{equation*}
F_{\textbf{1}}(x_{1},x_{2})=\frac{c}{|x_{1}-x_{2}|^{2h_{2,1}-h_{3,1}}}=\frac{c}
{|x_{1}-x_{2}|^{-2/\kappa}}\,.
\end{equation*}
{F}rom this expression for $F_{\textbf{1}}(x_{1},x_{2})$, it can be
seen that
$$
\frac{\p F_{\textbf{1}}(x_{1},x_{2})}{\p
x_{1}}=\frac{2}{\kappa}\frac{F_{\textbf{1}}(x_{1},x_{2})}{|x_{1}-x_{2}|}\,.
$$
Substituting this into Eq.~(\ref{reven}), we obtain
\begin{equation}\label{FP}
\Big[ \frac{\kappa}{2}\frac{\p}{\p
x_{1}^2}+\frac{2}{x_{1}-x_{2}}(\frac{\p}{\p x_{2}}-\frac{\p}{\p
x_{1}}) \Big]G(x_{1},x_{2},\Phi)=\delta_{x_{1}}
G(x_{1},x_{2},\Phi)\,.
\end{equation}
This has the form of the adjoint Fokker-Planck equation corresponding to the stochastic process
\begin{align}
dx_{1}&=\sqrt{\kappa}dB_{t}+\frac{2}{x_{1}-x_{2}}dt\nonumber\\
dx_{2}&=\frac{2}{x_{2}-x_{1}}dt\,.\label{sto}
\end{align}
This is called SLE$(\kappa,2)$ \cite{CardyMult}. Since the differential equation resulting
from this choice of stochastic variables is the same as the CFT solution to the 2
curve problem, we conjecture that this stochastic process gives the driving
term in the Loewner equation for one curve, given the existence of the other.
Recently Dub\'edat\cite{DubedatPR} has argued that such a
description follows from the requirement that the generators of
the Loewner processes for the two curves should commute.

It is interesting to take $\kappa=0$, which results in a deterministic Loewner equation with
analytic solution. Solving Eq.~(\ref{sto}) yields the following forcing function
\begin{equation}
x_{1}=\sqrt{2t+\frac{\delta^2}{4}}\,,
\end{equation}
where $t$ parametrises distance along the curve and $\delta$ is
the distance between the starting points of the curves on the real
axis. Kadanoff et al. derived solutions to the Loewner equation for various driving terms
\cite{KaNiKa}, including $x_{1}\propto\sqrt{t}$. The solution below is similar to the case
of a square root forcing term. The mapping from the upper half plane with boundary curves
grown up to time $t$ back to the upper half plane $g_{t}$ is the
solution to the Loewner equation
\begin{equation}
\frac{\dd g_{t}(z)}{\dd
t}=\frac{2}{g_{t}(z)-\sqrt{2t+\frac{\delta^2}{4}}}
\end{equation}
subject to the boundary condition $g_{t=0}(z)=z$. A change of
variables to
\begin{equation*}
G_{t}=\frac{g_{t}}{\sqrt{\frac{\delta^2}{8}+t}}\,,\quad
\tau=\ln(\frac{\delta^2}{8}+t)\,,
\end{equation*}
leads to the equation
\begin{equation}
\frac{\dd G}{\dd \tau}=\frac{(G-y_{+})(G-y_{-})}{2(\sqrt{2}-G)}\,,
\end{equation}
with $y_{+}=2\sqrt{2}$ and $y_{-}=-\sqrt{2}$. In terms of
\begin{equation}
H(G)=\frac{4\sqrt{2}\ln(G+\sqrt{2})+2\sqrt{2}\ln(G-2\sqrt{2})}{3\sqrt{2}}\,,
\end{equation}
the equation may be written
$$
\frac{\dd H}{\dd \tau}=-1\,,
$$
with solution
$$
-H(\frac{g_{t}}{\sqrt{t+\frac{\delta^2}{8}}})=\ln(\frac{\delta^2}{8}+t)+\textrm{const}\,.
$$
The constant may be set by the requirement that $g_{t=0}(z)=z$,
then
\begin{equation}
-H(\frac{g_{t}}{\sqrt{t+\frac{\delta^2}{8}}})=\ln(1+\frac{8t}{\delta^2})-H(\frac{2\sqrt{2}z}{\delta})\,.
\end{equation}
The boundary curves $z_{c}(t)$ are the line of
singularities which are found by setting
$g_{t}(z)=\sqrt{2}\sqrt{t+\delta^2/8}$, namely
\begin{equation}
-H(\sqrt{2})=\ln(1+\frac{8t}{\delta^2})-H(\frac{2\sqrt{2}z_{c}(t)}{\delta})\,.
\end{equation}
After substitution for $H(G)$ this may be simplified to
$$
4(\frac{z_{c}(t)}{\delta})^3-3\frac{z_{c}(t)}{\delta}+f(t)=0\,,
$$
with $f(t)=2(1+8t/\delta)^{3/2}-1$. The solutions are hyperbole of
the form
\begin{equation}
4a^2-\frac{4}{3} b^2=\delta^2\,,
\end{equation}
where the location of the tip is given by $z_{c}=a+ib$. The limit
$\delta\rightarrow 0$ is the case which we have quoted in
Sec.~\ref{kappa0}. In this limit, the curve is a straight line,
proceeding at an angle of $\pi/3$ from the positive real axis. By
symmetry, the other curve is also a straight line, making an angle
of $\pi/3$ from the negative real axis. Figure~(\ref{kzero})
displays the solution.

\section{Summary}
This paper has described how the conformal field theoretic treatment of SLE may be generalised to two curves.
In the limit that both curves originate from the same point, the equations for the probability that a
point lies to the left, right or between the two curves simplify
to a third order ordinary differential equation. This is the limit which has been investigated in this paper.
Results have been obtained for the range $0\leq\kappa\leq 8$ in
terms of integrals of hypergeometric functions. The special cases of $\kappa=0,2,4,8/3,8$ allow exact analytic
solutions in terms of elementary functions.

The application of the result for $\kappa=6$ to the quantum Hall problem has been explained, along with
its limitations.

It would be interesting to investigate the generalisation of the work in this paper to the case of $n$ curves
starting from the origin.
Although no more difficult in principle, the mathematics would be complicated; the solutions
are those of $(n+1)$th order ordinary differential equations.

\em Acknowledgments\em:

This work was supported in part by EPSRC Grant
GR/R83712/1. AG was supported by an EPSRC Studentship. The authors are grateful
to John Chalker for helpful discussions.

\appendix
\section{Level 3 null states}\label{l3n}
The Virasoro generators have the following commutation relations:
$$
[L_{m},L_{n}]=(m-n)L_{m+n}+\frac{c}{12}m(m^2-1)\delta_{m+n}\,.
$$
The aim of this section is to find the conditions for an operator to have a
null state at level 3, which is to say that
\begin{equation}\label{l3nsc}
L_{n}(L_{-3}+\mu L_{-2}L_{-1}+\lambda L_{-1}^{3})|\phi\rangle=0\,.
\end{equation}
Choosing $n=1,2$ in the equation above, leads to equations which
determine $\mu$, $\lambda$ and $h_{3}$, the scaling dimension of
$\phi$, as functions of of the central charge, $c$. First, acting
with $L_{1}$, dropping the $|\phi\rangle$ for clarity:
\begin{align*}
0&=L_{1}L_{-3}+\mu L_{1}L_{-2}L_{-1}+\lambda
L_{1}L_{-1}^3\\
&=(4+2\mu h_{3})L_{-2}+(3\mu+6\lambda+6\lambda
h_{3})L_{-1}^{2}\,.
\end{align*}
The coefficients of $L_{-2}$ and $L_{-1}^{2}$ must both vanish,
since otherwise this would imply a null state at level 2. Hence
\begin{align*}
4+2\mu h_{3}=0\\
3\mu+6\lambda(1+h_{3})=0\,.
\end{align*}
These simultaneous equations have solutions:
\begin{equation*}
\mu=-\frac{2}{h_{3}}\,,\quad
\lambda=\frac{1}{h_{3}(1+h_{3})}\,.
\end{equation*}
Consider Eq.~(\ref{l3nsc}) with $n=2$ to obtain the dependence of
$h_{3}$ on the central charge, c:
\begin{align*}
0&=L_{2}L_{-3}+\mu L_{2}L_{-2}L_{-1}+\lambda
L_{2}L_{-1}^3\\
&=(5+\frac{\mu c}{2}+4\mu h_{3}+4\mu+18\lambda
h_{3}+6\lambda)L_{-1}\,.
\end{align*}
Hence,
$$
5+\mu(\frac{c}{2}+4h_{3}+4)+\lambda(18h_{3}+6)=0\,.
$$
Substituting for $\mu$ and $\lambda$ from above,
$$
5-\frac{2}{h_{3}}(\frac{c}{2}+4h_{3}+4)+\frac{1}{h_{3}(1+h_{3})}(18h_{3}+6)=0\,,
$$
which is the following quadratic equation in $h_{3}$:
$$
3h_{3}^{2}+h_{3}(c-7)+c+2=0\,.
$$
This has solution
\begin{equation}\label{delta}
h_{3}=\frac{7-c\pm\sqrt{(c-7)^{2}-12(c+2)}}{6}\,.
\end{equation}
Hence, $\mu$, $\lambda$ and $h_{3}$ are all restricted to given
functions of $c$, the central charge, or equivalently in terms of
$\kappa$, the \sle variable:
$$
h_{3}=\frac{8-\kappa}{\kappa}\,.
$$
\section{From the correlation function to the differential equation}\label{Lminus3}
Consider the correlation function $P$ defined by:
\begin{displaymath}
  P=\lim_{y\rightarrow\infty}\frac{\langle\Phi(\xi)\phi_{3}
  (x_{1})\phi_{2}^2(y)\rangle}{\langle\phi_{3}(x_{1})\phi_{2}^2(y)\rangle}\,.
\end{displaymath}
The contour integral associated with the raising operator $L_{-3}$
acting on the state at $x_{1}$ can be deformed continuously until
it surrounds $\xi$. Hence
\begin{displaymath}
0=\langle\epsilon
L_{-3}\phi_{3}(x_{1})\Phi(\xi)\phi_{2}^2(\infty)\rangle+
\langle\phi_{3}(x_{1})\Phi(\xi+\frac{\epsilon}{(\xi-x_{1})^{2}})\phi_{2}^2(\infty)\rangle\,.
\end{displaymath}
Using the level three null state condition (see Appendix~\ref{l3n}),
\begin{displaymath}
  (L_{-3}+\mu L_{-2}L_{-1}+\lambda L_{-1}^{3})\phi_{3}=0\,,
\end{displaymath}
this equation may be written as
\begin{equation}\label{corr}
0=-\langle\epsilon\mu
L_{-2}L_{-1}\phi_{3}\Phi\rangle-\langle\epsilon\lambda
L_{-1}^{3}\phi_{3}\Phi\rangle+\epsilon2{\tr Re}\Big[\frac{1}{(\xi-x_{1})^{2}}\frac{\dd}{\dd\xi}\Big]P\,,
\end{equation}
where the operators at infinity and the operators' dependence on
position have been dropped for clarity. The first term, involving
$L_{-2}L_{-1}$, can be re-written as
\begin{displaymath}
  \langle\mu L_{-1}\phi_{3}\oint\frac{T(z)}{z-x_{1}}\frac{dz}{2\pi
  i}\Phi(\xi)\rangle\,,
\end{displaymath}
where a cancelling minus sign has appeared from reversing the
direction of the contour from clockwise to counter-clockwise. The
integral is equivalent to $\Phi(\xi)\rightarrow\Phi(\xi')$, using
that the scaling dimension of $\Phi$ is zero and defining $\xi'$
as
\begin{displaymath}
  \xi'=\xi+\frac{\epsilon}{\xi-x_{1}}\,.
\end{displaymath}
Writing $\xi$ in terms of real and imaginary parts as
$\xi=u+\textit{i}\ v$,
\begin{displaymath}
  \delta\xi=\epsilon\frac{(u-x_{1})-v\
  \textit{i}}{(u-x_{1})^{2}+v^{2}}\,.
\end{displaymath}
Then, the contribution to Eq.~(\ref{corr}) is
\begin{displaymath}
  \epsilon\mu\frac{\partial}{\partial
  x_{1}}\Big(\frac{u}{u^{2}+v^{2}}\frac{\partial}{\partial u}-\frac{v}{u^{2}+v^{2}}\frac{\partial}{\partial v}\Big)
  P(u-x_{1})\,.
\end{displaymath}
The contribution from the second term in Eq.~(\ref{corr}) may
be written as
\begin{equation*}
  -\epsilon\lambda\langle L_{-1}^{3}\phi_{3}\Phi\phi_{2}^{2}\rangle=-\epsilon\lambda\frac{\partial^{3}}{\partial
  x_{1}^{3}}P(u-x_{1})
  =\epsilon\lambda\frac{\partial^{3}}{\partial
  u^{3}}P(u-x_{1})\,,
\end{equation*}
where we have used that the function $P(x_{1},u,v)$ must be a
function of $P(u-x_{1},v)$, ie.
\begin{displaymath}
\frac{\partial P(u-x_{1})}{\partial x_{1}}=-\frac{\partial
P(u-x_{1})}{\partial u}\,.
\end{displaymath}
Lastly, we consider the contribution from the third term in
Eq.~(\ref{corr}). Again writing $\xi=u+\textit{i}v$,
\begin{align*}
\delta\xi&=\frac{\epsilon}{(u-x_{1}+\textit{i}v)^{2}}\\
&=\epsilon\frac{u^{2}-v^{2}}{(u^{2}-v^{2})^{2}+4u^{2}v^{2}}-
\epsilon\frac{2uv\textit{i}}{(u^{2}-v^{2})^{2}+4u^{2}v^{2}}\,,
\end{align*}
where $x_{1}$ has been set as the origin. The
contribution to Eq.~(\ref{corr}) is
\begin{align*}
&\epsilon\frac{u^{2}-v^{2}}{(u^{2}-v^{2})^{2}+4u^{2}v^{2}}\frac{\partial}{\partial
u}-\epsilon\frac{2uv}{(u^{2}-v^{2})^{2}+4u^{2}v^{2}}\frac{\partial}{\partial
v}\\
&=\epsilon\frac{u^{2}-v^{2}}{(u^{2}+v^{2})^{2}}\frac{\partial}{\partial
u}-\epsilon\frac{2uv}{(u^{2}+v^{2})^{2}}\frac{\partial}{\partial
v}\,.
\end{align*}
Putting all three terms together and using $P=f(u-x_{1})$,
\begin{equation*}
\Big[-\frac{\mu}{u^{2}+v^{2}}\Big(u\frac{\partial}{\partial
u}-v\frac{\partial}{\partial
  v}\Big)\frac{\partial}{\partial
  u}+\lambda\frac{\partial^{3}}{\partial
  u^{3}}
  +\frac{(u^{2}-v^{2})\frac{\partial}{\partial
u}-2uv\frac{\partial}{\partial
v}}{(u^{2}+v^{2})^{2}}\Big]P(\frac{u}{v})=0\,.
\end{equation*}
Next make the following substitutions:
\begin{itemize}
  \item $\frac{\partial P(t)}{\partial u}=\frac{1}{v}\frac{dP(t)}{dt}$
  \item $\frac{\partial P(t)}{\partial v}=-\frac{t}{v}\frac{dP(t)}{dt}$\,.
\end{itemize}
Then the differential equation can be rewritten as
\begin{align*}
&\Big[ -\frac{u\mu/v^2}{u^2+v^2}\frac{d^2}{dt^2}+
\frac{v\mu}{u^2+v^2}\Big(\frac{-t}{v^2}\frac{d^2}{dt^2}-\frac{1}{v^2}\frac{d}{dt}\Big)+\frac{\lambda}{v^3}\frac{d^3}{dt^3}\\
&\qquad+\frac{u^2-v^2}{(u^2+v^2)^{2}}\frac{1}{v}\frac{d}{dt}
+\frac{2uv}{(u^2+v^2)^2}\frac{u}{v^2}\frac{d}{dt}\Big]P(t)=0\,.
\end{align*}
Multiplying both sides by $v^3$ the ordinary differential equation becomes:
\begin{equation*}
\Big[ -2\frac{\mu
t}{t^2+1}\frac{d^2}{dt^2}-\frac{\mu}{t^2+1}\frac{d}{dt}+\lambda\frac{d^3}{dt^3}+\frac{t^2-1}{(t^2+1)^2}
\frac{d}{dt}+\frac{2t^2}{(t^2+1)^2}
\frac{d}{dt}\Big]P(t)=0\,.
\end{equation*}
Collecting terms, this is:
\begin{displaymath}
\lambda\frac{d^{3} P}{dt^{3}}-\frac{2\mu
t}{t^{2}+1}\frac{d^{2}P}{dt^{2}}+\frac{(3-\mu)t^2-(1+\mu)}{(t^2+1)^{2}}\frac{dP}{dt}=0\,,
\end{displaymath}
which is Eq.~(\ref{ode}) in the text.
\newpage

\end{document}